\documentclass[aps,titlepage,12pt,superscriptaddress]{revtex4}
\usepackage{epsfig}
\usepackage{times}
\usepackage{color}
\begin{document}

\title{Individual wealth-based selection supports cooperation in spatial public goods games}

\author{Xiaojie Chen}
\email{xiaojiechen@uestc.edu.cn}
\affiliation{School of Mathematical Sciences, University of Electronic Science and Technology of China, Chengdu 611731, China}

\author{Attila Szolnoki}
\email{szolnoki@mfa.kfki.hu}
\affiliation{Institute of Technical Physics and Materials Science, Centre for Energy Research, Hungarian Academy of Sciences, P.O. Box 49, H-1525 Budapest, Hungary}

\begin{abstract}\noindent
\\ In a social dilemma game group members are allowed to decide if they contribute to the joint venture or not. As a consequence, defectors, who do not invest but only enjoy the mutual benefit, prevail and the system evolves onto the tragedy of the common state. This unfortunate scenario can be avoided if participation is not obligatory but only happens with a given probability. But what if we also consider a player's individual wealth when to decide about participation? To address this issue we propose a model in which the probabilistic participation in the public goods game is combined with a conditional investment mode that is based on individual wealth: if a player's wealth exceeds a threshold value then it is qualified and can participate in the joint venture. Otherwise, the participation is forbidden in the investment interactions. We show that if only probabilistic participation is considered, spatially structured populations cannot support cooperation better than well-mixed populations where full defection state can also be avoided for small participation probabilities. By adding the wealth-based criterion of participation, however, structured populations are capable to augment network reciprocity relevantly and allow cooperator strategy to dominate in a broader parameter
interval.

\end{abstract}


\maketitle

\noindent

Cooperation is imperative when humans deal with problems of collective action, such as global warming, overpopulation,
overfishing or preserving natural resources \cite{hardin_g_s68,ostrom_90}. The target is always clear but the emerging temptation is also apparent. How can cooperation evolve in a population of rational individuals who pose a puzzle: why should cooperators incur a cost to benefit others \cite{pennisi_s05}? Evolutionary game theory is a powerful framework to study the problem of cooperation and many efforts have been considered in the past years to clarify the raised problem \cite{szabo_pr07,perc_bs10,nowak_jtb12,rand_tcs13,perc_jrsi13,li_pre15,suzuki_pre08,tian_pa16,wu_jrsi16}.

One of the potentially promising avenues which can address the irreconcilable conflict of individual and collective benefits is when we consider personal wealth of players which is already the product of their past interactions. Recent studies demonstrated that wealth heterogeneity and accumulation can promote cooperation \cite{wang_j_pre10b,perc_pre11,chadefaux_pone10,tavoni_pnas11,milinski_cc11,vasconcelos_pnas14,abou_chakar_jtb14,perc_njp11},
but making wealth visible hinders the evolution of cooperation \cite{nishi_n15,cote_pnas15,li_srep16}. Although in the scenario of wealth heterogeneity the richer players generally contributes more than the poorer partner, but high economic inequality leads higher-income individuals to be less generous \cite{cote_pnas15}. In addition, voluntary participation has offered an escape route to avoid the tragedy of the common state \cite{hauert_s02}.
Motivated by the latter option, Sasaki {\it et al.} investigated the evolutionary dynamics in the well-mixed public goods game with probabilistic participation rule \cite{sasaki_prsb07}. Similar probabilistic participation in the prisoner's dilemma game is found to favor the evolution of cooperation both in well-mixed and spatially structured populations \cite{traulsen_jtb07,chen_xj_pre08b}. Nevertheless previous works assumed that individuals have the initiative to decide whether or not to participate in the joint venture. There are cases, however, when players should fulfill certain criterion for getting the chance to participate. We may quote that applying for a club-membership or other examples when a closed community demands a qualification process from newcomers. An assessment factor for qualification could be individual wealth. If the wealth is large enough, an applicant should be qualified who then decides whether or not to participate. Even if a successful qualification some may not participate in the investment club because participation may not necessarily be profitable when the group is occupied by cheaters or the participation fee is too high \cite{masuda_prsb07,sasaki_pnas12}.

Inspired by these real-life experiences in the present work we combine individual wealth with the probabilistic participation into the public goods game. Our principal goal is to clarify whether such conditional participation mode is able to promote cooperation better than the pure probabilistic participation mode mentioned earlier. We assume that each individual has a varying wealth in the population. If the individual's wealth exceeds a specific threshold value then the player is offered the opportunity to participate in the group effort. In the other case, when personal wealth is below the threshold then the player is disqualified and cannot participate in the joint venture. In agreement with previous works the qualified player's choice is characterized by a probability factor. In case of positive decision the player pays a participation cost independently of his own strategy. Furthermore, a cooperator will also contribute to the common pool which will be enhanced and redistributed among all group members. Consequently, the individual's wealth is updated based on the collected payoff. As we will show, spatial structure cannot promote the evolution of cooperation better than well-mixed interactions when only probabilistic participation rule is applied. The combination with the wealth-based investment mode, however, is capable to provide a feedback which can amplify the positive consequence of network reciprocity, hence the advantage of structured population is revealed.

\section*{Results}
\noindent We first present the fraction of cooperators in the equilibrium state as a function of the participation cost $g$ and the participation probability $p$. Figure~\ref{fig1}(a) illustrates that when the participation probability is less than $0.6$, full cooperation state is always achieved independently of the participation cost value, $g$. By using an intermediate value of participation probability, i.e., when $0.6 < p < 0.9$, the fraction of cooperators first increases from zero until reaching a maximum value and then decreases as we increase the participation cost $g$. If we increase $p$ further the system always terminates onto a full defection state regardless of the cost value $g$. Figure~\ref{fig1}(b) shows the fraction of qualified individuals, whose individual wealth exceeds the threshold level $W_T$ in the equilibrium state. This panel highlights that all individuals are
qualified to participate in the investment game if $p$ is less than $0.4$ no matter how high participation cost in involved. This observation supports a previous finding, namely rare interactions reveal the positive consequence of cooperator strategy \cite{wang_z_srep12}. For an intermediate values of participation probability, when $0.4<p<0.6$, the actual value of $g$ becomes decisive: if the participation cost is small, all individuals are qualified, but their fraction decreases gradually as we increase $g$. Increasing $p$ further the role of $g$ becomes irrelevant again. It is because too frequent interactions provide an easy target for defectors who prevail and pull down the average individual wealth. As a result, all players become disqualified during the evolutionary process. Our last panel in Fig.~\ref{fig1}(c) suggests that there is a strong correlation between the fraction of cooperators and qualified players. A player's wealth can only be maintained if he cooperates with other group members otherwise the failure of group product will reflect on individual success. In this way the wealth-based investment rule can
augment the network reciprocity  mechanism effectively.

In order to identify the pure consequence of wealth-based investment mode on the evolution we now investigate the evolutionary dynamics of cooperation in well-mixed population without involving the wealth factor, that is, applying formally $W_T=-\infty$. According to the replicator equation, we find that there exists a critical participation probability $p^*$ above which the state of full cooperation is the only stable state in the system, that is, cooperators dominate the whole population (see Methods). While for $p<p^*$, the system always terminates onto the full defection state. We find that the $p^*$ critical value depends both on the value of the group size $N$ and the enhancement factor $r$. More precisely, $p^*$ decreases with increasing group size $N$, while it increases with increasing the enhancement factor $r$. This behavior is summarized in Fig.~\ref{fig2}(a). For the sake of comparison, we also plot in Fig.~\ref{fig2}(b) the equilibrium fraction of cooperators for $W_T=-\infty$ in spatially structured populations by using the parameter settings of Fig.~\ref{fig1}. The right panel in Fig.~\ref{fig2} highlights that cooperators cannot survive in spatial structure for $p>0.6$ practically independently of the participation cost value, but they dominate the whole system for $p<0.6$. We note that the critical value $p^*$ is approximately equal to $0.6$ that is the critical value of well-mixed population for the same parameter values [Fig.~\ref{fig2}(a)]. These results indicate that spatially structured population cannot facilitate the evolution of cooperation significantly, comparing with well-mixed population if only probabilistic participation mode is applied. The
comparison of Fig.~\ref{fig1}(b) and Fig.~\ref{fig2}(b), however, clearly demonstrates that the combination of wealth-based investment with probabilistic participation can provide cooperation supporting environment in structured population.

To get a deeper understanding about the microscopic process that is responsible for the cooperator supposing mechanism in spatially structured populations, we present a series of snapshots of strategy evolution for three representative participation cost values in Fig.~\ref{fig3}. When producing the snapshots we use different colors not just to distinguish cooperator and defector strategies, but also to mark their wealth status which determines if they are qualified to participate in the game or not. More precisely, blue (yellow) color denotes cooperators (defectors) whose wealth exceeds the $W_T$ threshold level, while green (red) color denotes cooperators (defectors) whose wealth is below the threshold level hence they are disqualified from the game.

For small participation cost, as shown in the top row of Fig.~\ref{fig3}, both cooperators and defectors remain qualified to participate in the games which offers an easy prey for defectors. Because of the small cost cooperators do not turn to ``green state" which would reduce their exploitation by defectors. As a result, blue cooperators vanish and yellow defectors prevail. However, these ``yellow" defectors' wealth decreases gradually because they are
unable to exploit others anymore and instead they still have to pay the participation cost. Consequently, the whole system turns into ``red state" at the end.

For intermediate participation cost, illustrated in the middle row of Fig.~\ref{fig3}, defectors can exploit cooperators at the beginning of the evolution, but they still turn to ``red" state quickly due to the increased participation cost. Cooperators are facing to the same problem therefore they also switch from ``blue" to ``red" state. Only those cooperators can maintain their qualifications who are deep in the middle of a surviving cooperator
domain because the enhanced benefit of public goods game can compensate the relative high value of $g$. Interestingly, they remain protected from being exploited by defectors, because a shield of ``green" cooperators will prevent defector players to contact directly with qualified, hence vulnerable, cooperators. In other words, the failure of cooperators in the vicinity of defectors will block the propagation of defector strategy. Finally, blue and green cooperators coexist in the population.

For large participation cost, as shown in the bottom row of Fig.~\ref{fig3}, defectors are capable to beat cooperators at the early stage of evolution because there are no organized cooperator clusters due to random initial condition. As earlier, large homogeneous defector cluster become disqualified (they change from yellow to red) because they cannot compensate the large participation fee from the income of games. Similar erosion can be observed for blue cooperators because they are unable to maintain the threshold level of wealth even if the absence of defectors in the neighborhood. As a result, both defectors and cooperators become disqualified and a voter-model-like neutral drift starts between the red and green domains. This slow evolution will eventually result in a homogeneous state where the probability of arriving at one of the possible final destinations is proportional to the initial portions of the mentioned strategies \cite{cox_ap83,dornic_prl01}.

To explore the robustness of our findings we have applied several different values of wealth threshold $W_T$. As Fig.~ \ref{fig4}(a) indicates clearly the non-monotonous dependence of cooperation level on the participation cost $g$ can still be observed for a wide range of $W_T$ interval. Correspondingly, full cooperation state can always be reached at intermediate values of the participation cost $g$. The only role of the value of $W_T$ is when to reach the
full-$C$ state. As the first panel of Fig.~ \ref{fig4} shows, lower threshold value requires a higher cost to reveal the collective benefit of cooperator strategy. In the middle panel we have plotted the fraction of qualified players in dependence of participation cost for different $W_T$ values. Figure~\ref{fig4}(b) shows that the number of qualified players is maximized at an intermediate $g$ value. This maximum is more pronounced as we increase the $W_T$
value. Figure~\ref{fig4}(c), where we plotted the fraction of qualified cooperators, suggests that our previous conclusion about the strong correlation of cooperative state and qualification is broadly valid, which explains why wealth-based selection rule augments the positive consequence of network reciprocity.

To complete our study we also investigated the fraction of strategies in dependence of participation cost for different initial wealth endowments, $W_0$. Our observations are summarized in Fig.~\ref{fig5}. We find that the non-monotonous dependence of cooperation level on the participation cost $g$ can still be observed for wide variety of $W_0$. The only role of $W_0$ is that it may cover the destructive consequence of defection: if $W_0$ is too high then all players are qualified in the system hence we get back the original public goods game where network reciprocity can support cooperation only in a limited way. If, however, the initial endowment of players is low then positive feedback of wealth-based selection manifests immediately providing a significantly higher cooperation level. This argument is justified by Fig.~\ref{fig5}(b) and (c) where the most striking change can be observed at small $W_0$ values. In other words, if the initial endowment of players is too high then we have no chance to observe the sophisticated mechanism we described above.

\section*{Discussion}
\noindent In public goods games it is a generally used assumption
that all players have a chance to participate in the joint venture.
But this postulation cannot be always justified. There are examples
when a certain criterion should be fulfilled by the applicants
before entering the investment club. Motivated by this real-life
experience we have proposed a so-called wealth-based selection rule
in the framework of spatial public goods game and combined it with
the previously studied probabilistic participation rule. Indeed,
probabilistic participation has been considered and studied in
well-mixed and spatially structured populations both for two-player
and multi-player games by some previous works
\cite{chen_xj_pre08b,sasaki_prsb07}. It is found that probabilistic
participation can promote the evolution of cooperation both in
well-mixed and spatially structured populations. However, it is
worth mentioning that these works have considered probabilistic
participation in well-mixed populations or in structured populations
independently and have not compared the possible impact of topology.
Therefore it remained unclear whether population structures can play
a positive role on the evolution of cooperation if probabilistic
participation rule is applied. An intuitive answer to this question
would be that cooperation is better promoted in structured
populations because the latter gives a chance for spatial
reciprocity to work \cite{perc_jrsi13}. Surprisingly, our comparison
revealed that structured populations cannot properly enhance the
cooperation level in comparison to well-mixed populations. In
particular, the critical value of the participation probability
which separates the full-cooperation state from full-defection state
is very similar for both cases.

When we also apply wealth-based selection rule, however, the positive consequence of interaction topology becomes evident. In the latter case the full-cooperator state can be reached even at large participation probability if participation cost is appropriately chosen. Moreover, we have shown that using a large wealth threshold or a small starting wealth endowment can also boost the evolution of cooperation. Interestingly, the role of these parameters are conceptually similar because they help to establish cooperator supporting conditions which reveal the significantly different consequences of competing strategies. While defection is destructive hence a defector cannot maintain high wealth value on the expense of neighbors, a cooperator player is able to remain qualified by supporting group members mutually. This explains why this selection mechanism is capable to magnify the positive impact of network reciprocity.

As we have already emphasized, our wealth-based selection rule is motivated by real-life observations. Indeed, individual wealth has been considered in some previous works where individual wealth distribution was fixed \cite{santos_n08,kun_a_nc13,teng_pa14}. Here the heterogeneity of wealth distribution was proved to be beneficial to cooperation \cite{kun_a_nc13}, which fits conceptually to the positive impact of social interactions \cite{perc_pre08} or interaction graph heterogeneity \cite{santos_n08}. In our present work individual wealth is not just a varying quantity but also plays a selection criterion to judge whether an individual is qualified or not for participating in a game. The key message of present model study is wealth, as the result of individual success, should not only be the target of evolution, but also driving force of selection mechanism. This observation supports our general intuition that in long run success-driven coevolutionary rules are helpful for cooperation \cite{pacheco_ploscb06,szolnoki_epl08,helbing_pnas09,szolnoki_njp08,liu_yk_csf12,szolnoki_epl14b}.

\section*{Methods}
\subsection*{Model with wealth-based participation in spatially structured populations}

\noindent The game is staged on a $L\times L$ square lattice with periodic boundary conditions. Each player on site $x$ with von Neumann neighborhood is a member of five overlapping groups of size $N=5$, and it is initially designated either as a cooperator ($s_x=1$) or defector ($s_x=0$) with equal probability. During the interaction stage at time $t$, if each individual $x$' wealth $W_x(t)$ is less than the threshold $W_T$, individual $x$ is not allowed to participate in the public goods game. Otherwise, it tends to participate in the public goods game with a probability $p$. When individual $x$ participates in the public goods game, he has to pay a participation cost $g$ first independently of his strategy. Then, participating cooperators invest a fixed amount $c$ to the common pool, while participating
defectors invest nothing.

According to the previously established rules, when individual $x$ participates in the game centered by a player $i$, his payoff from the group $i$ is $\Pi_x^i=rN_{PC}c/N_P-s_x c-g$, where $N_P$ is the number of participants and $N_{PC}$ is the number of cooperators among the participants. Note that $N_{PC}\leq N_P\leq N$ because of the qualification rule. It is worth mentioning that when only one individual participates in the game (i.e., $N_P=1$), the player can still receive the investment return from the game if it is a cooperator. When all participants invest into the pool, each obtains a payoff $(r-1)c-g$, which is assumed to be positive value \cite{sasaki_pnas12}. Furthermore, without losing generality \cite{hauert_s02}, $c$ is set to one in this work. If a player $x$ is not allowed to participate in the game or does not participate in the game then his payoff is $\Pi_x^i=0$. Since an individual $x$ belongs to $N$ different groups his total payoff $\Pi_x$ is simply accumulated from all related $\Pi_x^i$ incomes.

After playing the games with all available neighbors, the individual wealth value is updated
\begin{equation}
W_x(t+1)=W_x(t)+\Pi_x,
\end{equation}
where $\Pi_x$ is the total payoff of player $x$ obtains at time $t$. In the beginning all players are given an initial endowment $W_x(0)=W_0$, which is higher than $W_T$. Otherwise the evolution would be trapped in a frozen state.

After each round, a player $x$ is given the opportunity to imitate the strategy of a randomly selected nearest neighbour $y$. The strategy transfer occurs with the probability
\begin{equation}
q=\frac{1}{{1 + \exp [({\Pi_x} - {\Pi_y})/K]}},
\end{equation}
where $K$ characterizes the uncertainty by strategy adoptions \cite{szabo_pre98}. Without losing generality \cite{szolnoki_pre09c}, we use $K=0.5$, so that it is very likely that better performing players will be followed, although those performing worse may occasionally be imitated as well.

As the key quantity, we measure the stationary fraction of cooperators $\rho_c=L^{-2}\sum_x s_x(\infty)$, where $s_x(\infty)$ denotes the strategy of player $x$ when the system reaches dynamical
equilibrium, i.e., when the average cooperation level becomes time-independent. Moreover, to get a better statistics the final outcome is averaged over $100$ independent runs.

\subsection*{Analysis for the case of qualified game in well-mixed populations}

For studying the evolutionary dynamics in infinite well-mixed populations without involving the wealth factor, we use the replicator equation \cite{Hofbauer_98}. Initially we assume that a fraction $x$ of the population is formed by cooperators while the remaining fraction $(1-x)$ are defectors. The related replicator
equation is
\begin{equation}
\dot{x}=x(1-x)(\Pi_C-\Pi_D),\label{eq1}
\end{equation}
where $\Pi_C$ and $\Pi_D$ are the average payoffs of cooperators and defectors, respectively. Next, let groups of $N$ individuals be sampled randomly from the population. The average payoff of cooperators is $\Pi_C=(1-p)\cdot0+p\cdot\Pi_C^P$, where the $\Pi_C^P$ payoff of participating cooperators is
\begin{equation}
\Pi_C^P =\sum_{i=0}^{N-1}\left(\begin{array}{c}
N-1\\i\end{array}\right)p^i(1-p)^{N-1-i}\sum_{j=0}^{i}\left(\begin{array}{c}
i\\j\end{array}\right)x^j(1-x)^{i-j}\bigg[ \frac{r(j+1)c}{i+1}-c-g \bigg].
\end{equation}
Similarly, the average payoff of defectors is $\Pi_D=(1-p)\cdot0+p\cdot\Pi_D^P$, where the $\Pi_D^P$ payoff of participating defectors is
\begin{equation}
\Pi_D^P =\sum_{i=0}^{N-1}\left(\begin{array}{c}
N-1\\i\end{array}\right)p^i(1-p)^{N-1-i}\sum_{j=0}^{i}\left(\begin{array}{c}
i\\j\end{array}\right)x^j(1-x)^{i-j}\bigg[\frac{rjc}{i+1}-g\bigg].
\end{equation}
With these definitions, the replicator equation has two boundary
equilibria, namely $x=0$ and $x=1$. On the other hand, interior
equilibria can be determined by the roots of the function
$g(x)=\Pi_C-\Pi_D$, thus obtaining
\begin{equation}
g(x)=\frac{rc}{N}\Big[1-(1-p)^N\Big]-pc.
\end{equation}
It follows that $g(x)$ is independent of $x$, and there is no interior equilibria in $(0, 1)$.

For the stability analysis of the replicator equation, we need to
know the sign of $g(x)$ function. To determine it we define the
continuous function $h(p)=g(x)$. It follows that $h(0)=0$ and
$h(1)=(r/N-1)c<0$. Moreover, $h'(p)=[r(1-p)^{N-1}-1]c$ with
$h'(0)=(r-1)c>0$ and $h''(p)=-(N-1)rc(1-p)^{N-2}<0$. We can thus
conclude that $h(p)$ is positive near $p=0$, and $h(p)$ has a unique
interior root $p^*$ in $(0, 1)$ with $h(p^*)=0$, which yields the
following conclusions about the stability analysis.

($1$) For $p\in (0, p^*)$, $h(p)>0$ so that $\Pi_C-\Pi_D>0$. As a result, $x=1$ is a stable equilibrium, while $x=0$ is an unstable equilibrium.

($2$) For $p\in (p^*, 1]$, $h(p)<0$ so that $\Pi_C-\Pi_D<0$. As a result, $x=0$ is a stable equilibrium, while $x=1$ is unstable equilibrium.

($3$) For $p=0$ or $p=p^*$, $h(p)=0$ so that $\Pi_C-\Pi_D=0$. As a result, the game becomes equivalent to neutral drift.

\clearpage

\noindent \textbf{Acknowledgments} \\
This research was supported by the National Natural Science
Foundation of China (Grants No. 61503062), by the Fundamental
Research Funds of the Central Universities of China, and by the
Hungarian National Research Fund (Grant K-120785).

\noindent \\ \textbf{Author contributions} \\
The authors designed and performed the research as well as wrote the paper.

\noindent \\ \textbf{Competing financial interests} \\
The authors declare no competing financial interests.

\clearpage

\begin{figure}
\caption{The stationary outcome of evolutionary dynamics in
spatially structured populations driven by wealth-based investment
mode. Panel~(a) shows the equilibrium fraction of cooperators in
dependence on the participation cost $g$ and the participation
probability $p$. Panel~(b) shows the fraction of qualified
individuals whose wealth exceeds the threshold value, while
panel~(c) shows the fraction of cooperators among qualified players.
Other parameters: $r=3.0$, $W_T=0$, $W_0=50$, and $L=100$.}
\label{fig1}
\end{figure}

\begin{figure}
\caption{Evolutionary dynamics obtained by probabilistic
participation in the absence of wealth-based investment rule.
Panel~(a) shows the critical participation probability $p^*$ below
which cooperators dominate the population in dependence on the group
size $N$ for different values of the enhancement factor $r$ in
well-mixed populations. Panel~(b) shows the fraction of cooperators
in dependence on the participation cost $g$ and the participation
probability $p$ in spatially structured populations. In the latter
case other parameters are $r=3.0$ and $L=100$.} \label{fig2}
\end{figure}

\begin{figure}
\caption{Evolution of spatial strategy distribution starting from a
random initial state for three representative values of
participation cost. Top row [from (a) to (d)] depicts the time
evolution for small cost ($g=0.1$), middle row [from (e) to (h)]
depicts the evolution for intermediate cost ($g=1$), while the
bottom row [from (i) to (l)] denotes the evolution for high cost
($g=2$). Cooperators (defectors) whose wealth exceeds the $W_T$
threshold value are denoted by blue (yellow), while cooperators
(defectors) whose wealth is below the threshold are denoted by green
(red). Other parameters are $p=0.7$, $r=3.0$, $W_0=50$, $W_T=0$, and
$L=100$.} \label{fig3}
\end{figure}

\begin{figure}
\caption{Panel~(a) depicts the equilibrium cooperation level as a
function of participation cost $g$ for different values of wealth
threshold $W_T$. Other parameters are $p=0.6$, $W_0=50$, and
$L=100$. The legend for $W_T$ values are in the middle panel.
Panel~(b) and (c) depict the fraction of qualified players and
qualified cooperators respectively by using the same parameter
values as stated for panel~(a).} \label{fig4}
\end{figure}

\begin{figure}
\caption{Panel~(a) depicts the equilibrium cooperation level as a
function of participation cost $g$ for different values of initial
wealth endowment $W_0$. Other parameters are $p=0.7$, $W_T=0$, and
$L=100$. The legend for $W_0$ values are in the middle panel.
Panel~(b) and (c) depict the fraction of qualified players and
qualified cooperators respectively by using the same parameter
values as stated for panel~(a).} \label{fig5}
\end{figure}

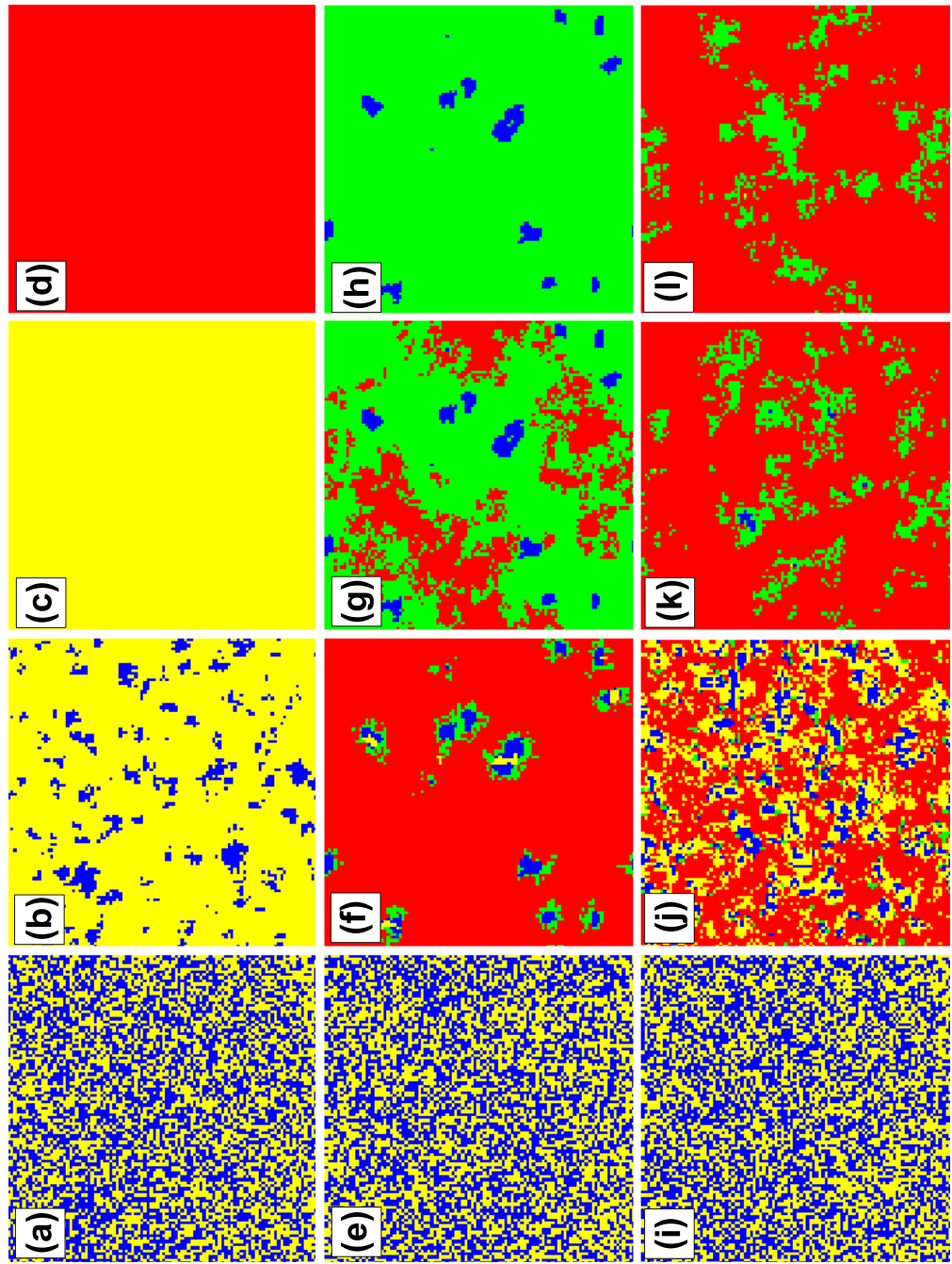
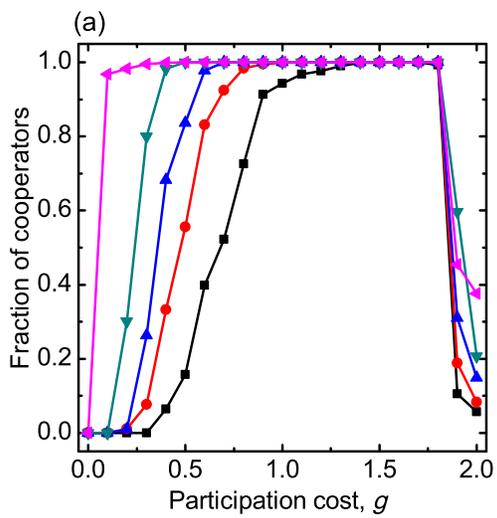
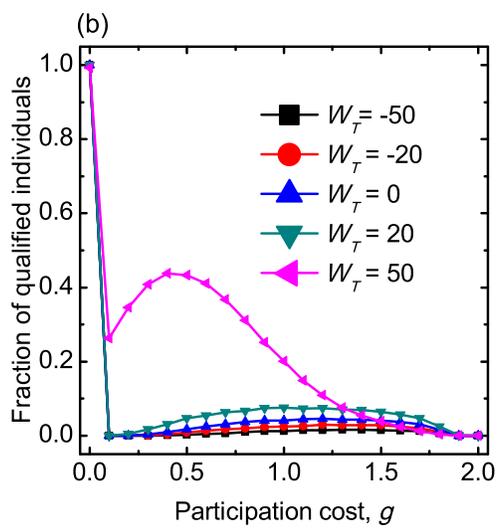
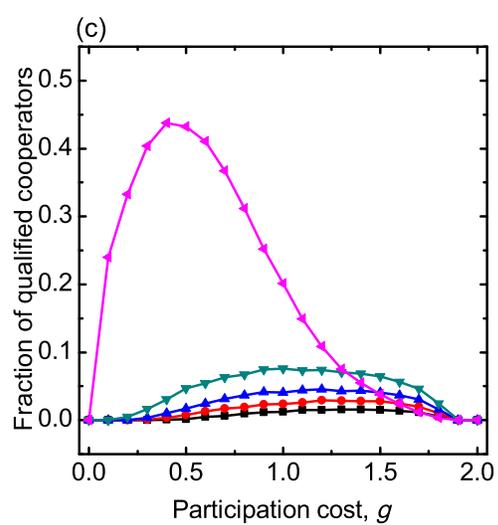
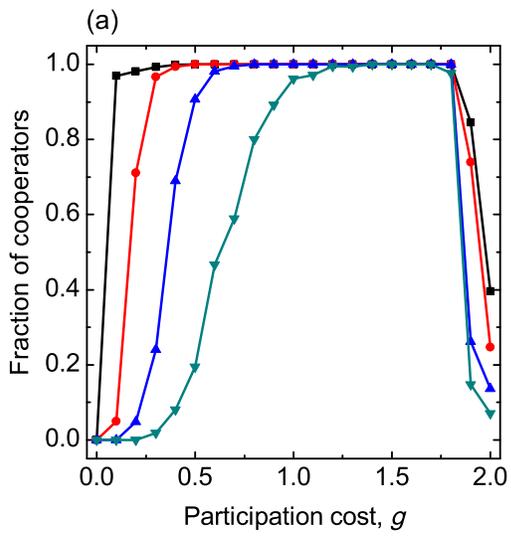
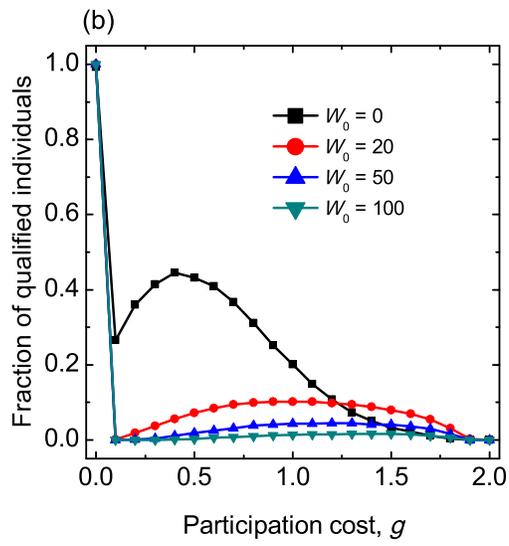
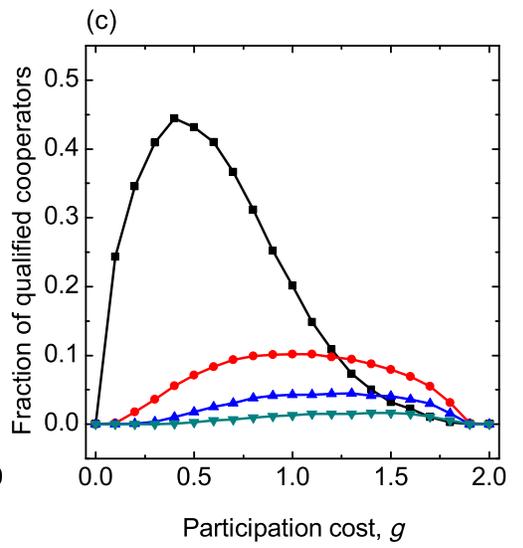
\end{document}